\documentclass[10pt, conference, compsocconf]{IEEEtran}

\usepackage{mdwmath}
\usepackage{mdwtab}
\usepackage{eqparbox}

\usepackage{stfloats}

\usepackage{adjustbox}
\usepackage{multirow}

\usepackage{wasysym}

\usepackage[caption=false,font=footnotesize]{subfig}

\usepackage{amsmath}
\usepackage{bbm}
\usepackage{stmaryrd}
\usepackage{amssymb}

\usepackage{array}
\usepackage{caption}
\usepackage[usenames, dvipsnames]{color}
\usepackage[hyphens]{url}
\usepackage[pdftitle=Title,pdfauthor=Anonymous]{hyperref}

\let\labelindent\relax
\usepackage{enumitem}

\usepackage{microtype}

\usepackage{xspace}
\newcommand{\etal}{\textit{et al.}\xspace}

\newcommand{\ie}{\textit{i.e.,}\xspace}
\newcommand{\eg}{\textit{e.g.,}\xspace}

\usepackage{todonotes}

\usepackage{draftwatermark}
\SetWatermarkColor[gray]{0.9}

\IEEEoverridecommandlockouts

\begin{document}
\title{SoK: Blockchain Technology and Its Potential Use Cases}

\makeatletter
\newcommand{\linebreakand}{%
	\end{@IEEEauthorhalign}
	\hfill\mbox{}\par
	\mbox{}\hfill\begin{@IEEEauthorhalign}
}
\makeatother

\makeatletter
\def\blfootnote{\gdef\@thefnmark{}\@footnotetext}
\makeatother

\author{
	\IEEEauthorblockN{Scott Ruoti}
	\IEEEauthorblockA{
		University of Tennessee, Knoxville\\
		MIT Lincoln Laboratory\IEEEauthorrefmark{1}\\
		ruoti@utk.edu
	}

	\and
	
	\IEEEauthorblockN{Ben Kaiser}
	\IEEEauthorblockA{
		Princeton University\\
		MIT Lincoln Laboratory\IEEEauthorrefmark{1}\\
		bkaiser@princeton.edu
	}

	\and
	
	\IEEEauthorblockN{Arkady Yerukhimovich}
	\IEEEauthorblockA{
		George Washington University\\
		MIT Lincoln Laboratory\IEEEauthorrefmark{1}\\
		arkady@gwu.edu
	}

	\linebreakand

	\IEEEauthorblockN{Jeremy Clark}
	\IEEEauthorblockA{
		Concordia University\\
		j.clark@concordia.ca
	}

	\and

	\IEEEauthorblockN{Robert Cunningham}
	\IEEEauthorblockA{
		Carnegie Mellon University\\
		Software Engineering Institute\\
		MIT Lincoln Laboratory\IEEEauthorrefmark{1}\\
		robertkcunningham@cmu.edu
	}	
}

\maketitle

\blfootnote{\IEEEauthorrefmark{1}The majority of this work was completed while these authors worked at MIT Lincoln Laboratory.}



\begin{abstract}
Bitcoin's success has led to significant interest in its underlying components, particularly Blockchain technology.
Over 10 years after Bitcoin's initial release, the community still suffers from 
a lack of clarity regarding what properties defines Blockchain technology, its 
relationship to similar technologies, and which of its proposed use-cases are 
tenable and which are little more than hype.
In this paper we answer four common questions regarding Blockchain technology: (1) what exactly is Blockchain technology, (2) what capabilities does it provide, and (3) what are good applications for Blockchain technology, and (4) how does it relate to other approache distributed technologies (e.g., distributed databases).
We accomplish this goal by using grounded theory---a structured approach to gathering and analyzing qualitative data---to thoroughly analyze a large corpus of literature on Blockchain technology.
This method enables us to answer the above questions while limiting researcher bias, separating thought leadership from peddled hype and identifying  open research questions related to Blockchain technology.
The audience for this paper is broad as it aims to help researchers in a 
variety of areas come to a better understanding of Blockchain technology and 
identify whether it may be of use in their own research.
\end{abstract}

\IEEEpeerreviewmaketitle
\thispagestyle{plain}
\pagestyle{plain}



\section{Introduction}

In 1982, David Chaum~\cite{Cha82} proposed using blind signatures to allow for untraceable electronic payments.
Later, Chaum, Fiat and Naor~\cite{chaum1988untraceable} expanded this idea into a fully fleshed out electronic cash system.
eCash was the first cryptocurrency---\ie ``a digital currency in which encryption techniques are used to regulate the generation of units of currency and verify the transfer of funds operating independently of a [nation-state-controlled] central bank.''\footnote{\url{https://en.oxforddictionaries.com/definition/cryptocurrency}}
However, early cryptocurrencies still required the presence of a central party to help manage the creation and transfer of funds.

Two decades after eCash was first proposed, an author using the pseudonym Satoshi Nakamoto wrote a white paper describing Bitcoin, a new decentralized cryptocurrency~\cite{Nak08}.
Unlike centralized cryptocurrencies which rely on a set of known entities to operate, Bitcoin uses a proof-of-work-based scheme~\cite{DN93,back1997partial} to allow the general public to maintain the system.
To incentivize public participation, Bitcoin pays participants (known as miners) for solving the proof-of-work puzzles.
While Bitcoin's building blocks were not novel, the composition of these properties into a single system was a meaningful contribution~\cite{Narayanan17} which led the cryptocurrency to become the first to achieve widespread popularity and usage.

At the peak of its popularity, Bitcoin's market capitalization reached \$835.69 billion (USD), though at the time of this writing it has seen a decrease in both its stature and market capitalization (\$60.78 billion).
Nevertheless, there remains significant interest in the question of whether the technology underlying Bitcoin---known as \emph{Blockchain technology} or {Blockchain} for short---could be used to build other interesting cryptocurrencies or distributed systems.
In particular, we have consistently heard researchers, business executives, and government leadership ask the following four questions: (1) what exactly is Blockchain technology, (2) what capabilities does it provide, and (3) what are good applications for Blockchain technology, and (4) how does it relate to other approache distributed technologies (e.g., distributed databases).

Unfortunately, there is a lack of clarity regarding the answers to these questions in both the community and the literature, impeding the ability for researchers to properly understand, evaluate, and leverage Blockchain technology.
This lack of clarity stems at least in part from competing viewpoints and terminology regarding Blockchain technology, resulting in seemingly contradictory statements that can all be true in their respective contexts.
In this paper, we attempt to address this ambiguity and provide a holistic overview of Blockchain technology, set forth a common lexicon for discussing Blockchain technology, and answer the questions identified above.

\begin{table}
	\centering
	
	\begin{tabular}{l|l}
		\textbf{Academia} & \textbf{Industry} \\

		Increased interest recently & Long-term interest \\
		Focused on specific subsystems & Takes a holistic view \\
		Technical experts & Use case experts \\ \hline
	\end{tabular}
	
	\caption{Academic vs. industrial sources}
	\label{tab:sources}
\end{table}

To accomplish these goals, we conducted a textual analysis of papers on Blockchain technology from non-academic sources (hereafter referred to as \emph{industry}).
The choice to focus our analysis on literature from industry was based on the unique suitability of that literature (see Table~\ref{tab:sources}) to answer the questions we identified above:
First, industry became heavily involved in Blockchain technology before academia and thus has had more time to fully consider its implications.
Second, industry literature tends to discuss and evaluate Blockchain technology holistically, which provides the broad perspective needed to answer questions one and two.
Finally, much of industry literature is written by use case experts who are well positioned to answer the third question.

In contrast, academic literature focuses on technical deep dives on Blockchain technology's subsystems and implementations built on Blockchain technology.
For example, existing academic surveys on Blockchain technology have focused on particular systems or technical properties: Bitcoin~\cite{BMC+15,Narayanan17}, payment privacy~\cite{Conti17}, security and performance~\cite{Gervais16}, scalability~\cite{Croman16}, and consensus protocols~\cite{Bano17,garay2018consensus}.
In this regard, the existing literature acts as a complement to this 
paper---i.e., researchers interested in Blockchain technology can read this 
paper to gain a holistic overview of the space and then dive deeply into 
specific subtopics by reviewing other Blockchain survey papers.



Ultimately, our technical analysis was successful at answering the questions 
identified previously, and in this paper we present those answers.
While some readers may find the results of this paper unsurprising or differing from their viewpoints regarding Blockchain technology, we believe that this paper will be helpful to the community at large (both Blockchain experts and non-experts) and provide a common reference point for discussion on Blockchain technology.
Taken together, the results in this paper represent the most complete overview of Blockchain technology and its potential use cases available in a single work that we are aware of.
The results are intended to serve as an aid for researchers as they field questions related to Blockchain and as they explore whether Blockchain technology is relevant to their personal research areas.
Concretely, our key contributions are:

\begin{enumerate}
	\item \textbf{Providing a holistic overview of Blockchain technology's technical properties and the capabilities they enable.}
	We organize the properties of Blockchain technology into three groups: shared governance and operation, verifiable state, and resilience to data loss.
	Taken together, these properties differentiate Blockchain technology from other distributed technologies.
	Using these properties, systems built with Blockchain technology have easy access to a range of important capabilities: full-system provenance, auditability, access control, psuedonymity, automatic execution (i.e., smart contracts), and data discoverability.
	
	
	
	\item \textbf{Identifying groups of applications (i.e., use cases) that are most likely to benefit from Blockchain technology.}
	Within the literature we analyzed there was a range of potential applications for Blockchain technology.
	We group these applications into a set of use cases and then discuss the likelihood that individual applications within the use case would benefit from the use of Blockchain technology.
	Example use cases include cryptocurrencies, asset management, and multi-organization data sharing.
	
	\item \textbf{Detailing challenges and limitations related to Blockchain technology.}
	As part of our review of the literature, we identified several important challenges and limitations for Blockchain technology: scalability, smart contract correctness and dispute resolution, stapling of on-chain tokens to off-chain assets, key management, and regulation.
	Many of these challenges represent important research questions with interesting potential for future research.


	\item \textbf{Leveraging grounded theory to analyze industrial literature while limiting research bias and separating hype from sound technical details.}
	While there are significant benefits to analyzing industrial literature, there is also a significant amount of hype and imprecise language.
	To address these limitations, we leveraged the grounded theory methodology~\cite{glaser1965constant,strauss1990basics,corbin1990grounded} (also known as the constant comparative method) to extract and separate valuable technical insights out from the hype and technical misunderstandings that permeate this body of work.
	Based on this analysis of data from industry, interspersed with our own knowledge and a review of the academic literature, we can answer the common questions we identified while remaining grounded in the data we analyzed.
	We were also able to shed light on industry's understandings and misunderstandings of Blockchain technology.

\end{enumerate}


\section{Methodology}
\label{sec:method}

To more holistically understand Blockchain technology, we conducted a rigorous textual analysis of Blockchain literature from non-academic sources (hereafter referred to as industry),\footnote{Here, we consider industry broadly: corporations, small and medium business, startups, and consortia.} including but not limited to source from the technology, financial, and healthcare industries.
While there is valuable information to be learned from these sources, analyzing them has certain challenges that while also present in academic literature are usually less pronounced:

\begin{enumerate}
	\item \emph{Lack of precise terminology and discussion.}
	In our review of materials from industry, we found that the same concepts were often described using divergent and imprecise terminology, leading to white papers that are difficult to understand and provide muddled descriptions of capabilities and use cases.
	Additionally, while there is a fair bit of factually inaccurate information in materials from industry (e.g., several documents claimed cryptographic signatures provide confidentiality), in several cases we  observed an accurate description of an idea that was phrased in such a way as to make it seem incorrect under cursory examination. Some of those ideas were assembled in ways that are interesting to academics, if one takes the time to work through the material.
	In this regard, materials from industry represent a trove of useful information obscured by imprecise terminology and discussion.
	
	\item \emph{Inclusion of hype.}
	Much of the material from industry includes visionary statements (i.e., hype) about how Blockchain technology will change business practices, power dynamics and the way the world works.
	This hype is a mixture of realistic use cases that can benefit from Blockchain technology (e.g., anonymous payments~\cite{chaum1988untraceable}) and ideals that far transcend any technical solution (e.g., removing the need for governments).
	Unfortunately, unlike what one would expect in academic literature, the materials from industry often intermingle hype with technical details. This at least partially explains why some in academia are quick to dismiss sources from industry.
	
	\item \emph{Researcher bias.}	
	Researcher bias is an obvious problem in any literature review---regardless of whether the source is academia or industry---and one that is often not explicitly addressed in systemization papers.
	The potential for bias is even stronger when reviewing materials from industry because the two issues described above (lack of precise terminology and hype) make it easy for researchers to dismiss out of hand ideas proposed by industry.
	
\end{enumerate}

For these three reasons, it is tempting to avoid analyzing materials from industry,
however, taking that approach would sacrifice significant insights into BLockchain technology and its use cases.
Instead we employ a well-established research method---\emph{grounded theory}~\cite{glaser1965constant,strauss1990basics,corbin1990grounded} (also known as the constant comparative method)---to rigorously analyze the data in a way that directly addresses each of these three problems.

Grounded theory is used to analyze qualitative data sources (e.g., user stories, interviews) and extract the underlying data and processes described across the myriad of gathered sources.\footnote{Grounded theory identifies data and processes that are supported across the body of sources and is not a method for creating a fine-grained breakdown of an individual document.}
In particular, grounded theory is designed to help researchers identify data and processes within qualitative data sources generated by humans and filled with imprecise terminology and descriptions.
Additionally, grounded theory limits the impact of researcher bias, ensuring that the data and processes are derived from the data and not from the researchers' preconceived notions of what the data says.
Grounded theory explicitly addresses the first and third problems we identified for evaluating materials from industry, and our hope was that it would also be able to separate the hype from the underlying data and processing. We believe these goals were satisfied based on our results. 

The idea of using grounded theory for literature review is not new~\cite{wolfswinkel2013using,yang2012descriptive} and this method has been used in thousands of studies examining qualitative data.\footnote{As evidence of its wide use, the top-cited paper describing grounded theory has 62,951 citations as of writing.}
For these reasons---and based on our own experience with the method~\cite{ruoti2017weighing}---we were confident this method would allow us to successfully accomplish our research goals.

In the remainder of this section we first describe how we gathered industry materials for our grounded theory analysis.
Next, we describe the grounded theory process in some detail, as it may be unfamiliar to readers in this field.
Lastly, we describe an academic literature review we conducted to enhance the results of our grounded theory analysis.

\subsection{Industry Material Gathering}
Beginning in the summer of 2016 we began to gather documents published regarding Blockchain technology.
This included both materials from industry and academia, though this section will focus on only the former.
We gathered materials using a variety of methods:

\begin{itemize}
	\item Following RSS feeds that track news and publications related to Blockchain technology (e.g., CoinDesk\footnote{\url{https://coindesk.com}}).
	\item Downloaded materials published by Blockchain consortiums (e.g., Hyperledger\footnote{\url{https://www.hyperledger.org/}}, Decentralized Identity Framework\footnote{\url{http://identity.foundation/}}) and their members (e.g., IBM, Microsoft, Gem).
	\item Using Google to explore what was being said about Blockchain technology by major accounting firms, banks, and tech companies.
	\item Browsing new articles and blog posts related to Blockchain technology. This included articles which gave lists of interesting Blockchain papers.
	\item Reviewing submissions to the ONC Blockchain in Health Care Competition.\footnote{\url{https://www.healthit.gov/topic/grants-contracts/announcing-blockchain-challenge}}.
\end{itemize}

When reviewing these materials, we would also follow references and include those documents if we believed they were relevant.
In total, we collected 132 documents across three categories.

\begin{itemize}
	\item \emph{High-Level Overviews.} These were often prepared by investment firms and gave high level overviews of Blockchain technology. They would also reference various efforts at using Blockchain in practice.
	\item \emph{System White Papers.} These papers would describe how Blockchain technology was used in a specific system, or more frequently a system proposal.
	\item \emph{Blockchain Commentaries.} These were generally shorter documents that would discuss a specific facet of Blockchain technology in greater depth than we saw in other documents.
\end{itemize}

\subsection{Grounded Theory Data Analysis}
After collecting our initial set of 104 documents, we analyzed them using grounded theory.
This methodology splits analysis of the documents into four stages: open coding, axial coding, selective coding, and theory generation.
Throughout the analysis of the documents we kept detailed research notes that outlined our thoughts as we reviewed and analyzed the literature.
Additionally, we conducted intensive discussion between the various researchers to ensure that we were correctly understanding and evaluating the source material.
As is often the case in grounded theory, these notes and discussion were every bit as important, if not more so, than the concepts, categories, and theories we generated.

\subsubsection{Stage 1---Open Coding}
In this first stage, documents were assigned to one of four reviewers.
Each reviewer would read the document, assign codes to words and sentences in the document.
These codes were generated using a mixture of open coding (assigning a code that summarizes the document's statement) and in situ coding (using the document's own words as the code).
To ensure that we were assigning the correct codes, we paid careful attention to the context of each statement.

In particular, reviewers made sure to code the following four concepts found in documents:
\begin{itemize}
	\item \emph{Properties.} What are the building blocks for Blockchain technology? What capabilities does it provide?
	\item \emph{Challenges.} What challenges must be addressed when building systems using Blockchain technology?
	\item \emph{Limitations.} What inherent limitations are there when using Blockchain technology?
	\item \emph{Use cases.} What applications or groups of applications (i.e., uses cases) benefit from the application of Blockchain technology?
\end{itemize}

At this stage of the grounded theory process, reviewers were instructed to avoid evaluating the validity of the coded concepts.
Instead, every attempt was made to include all possible codes, helping to ensure that our results were grounded in the data and not reviewers' biases.

The reviewers continued reviewing documents until each felt that the last 3--5 documents they had read had no concepts that had not already been brought up by previous documents.
This is a commonly accepted stopping criteria in grounded theory and is indicative that all core (i.e., not truly one-off) ideas have been discovered.
In total, this stage resulted in the creation of 641 codes.

\subsubsection{Stage 2---Axial Coding}
In the second stage, our research team used the constant comparative method to group codes into concepts.
Specifically, we collapsed distinct codes referring to the same topic (e.g., one was an open code, the other in situ) into a single code, reducing the original set of 641 codes to a more manageable 68 concepts.
As needed, we referred to the original documents to ensure that our understanding of the code was fresh, and that we were assigning it to the appropriate concept.
Also, at this stage we continued to avoid evaluating the validity of concepts, ensuring that the ideas of the reviewed documents were fully reflected in the codes.

\subsubsection{Interlude---Additional Open Coding.}
After completing axial coding, one reviewer coded (i.e., open coding) another 28 documents (giving the total of 132 documents).
These documents were all blog posts, representing the most up-to-date thinking on Blockchain technology.
In this process, no new codes were discovered, indicating that our process had produced concepts that thoroughly describe Blockchain technology.

\subsubsection{Stage 3---Selective Coding}
In the third stage, two researchers transferred the concepts related to technical properties and applications onto sticky-notes.
They then drew connecting lines between these concepts, describing how they related to one another.
Based on these interconnections, concepts were divided into five different categories:

\begin{itemize}
	\item \emph{Technical properties (Figures~\ref{fig:technical-properties},~\ref{fig:technical-properties-full}).}
	Technical properties are the components that make up of Blockchain technology. Examples include decentralized governance, a consensus protocol, and an append-only transaction ledger.
	
	\item \emph{Capabilities (Figure~\ref{fig:Capabilities}).}
	Capabilities are the high-level features provided by Blockchain technology's technical properties. Examples include automatic executions (i.e., smart contracts), internal auditability, and access control.
	
	\item \emph{Technical primitives.}
	Primitives are the building blocks used to construct the technical properties and capabilities of Blockchain technology. Examples include timestamps, hashchains, and peer-to-to-peer communication.
	
	\item \emph{Use cases.}
	Use cases are classes of systems that the literature identified as being good fits for Blockchain technology. Examples include crytocurrencies, supply chain management, and identity management.
	
	\item \emph{Normative properties (Figure~\ref{fig:normative-properties}).}
	Normative properties represent what people hope to achieve using Blockchain technology. 
	Importantly, these properties are not provided by the use of Blockchain technology---as are the technical properties and capabilities---but instead require the careful designs of larger systems that might only use Blockchain technology as a small piece of the overall system.
	In general, normative properties strongly relate to the hype surrounding Blockchain technology.
	Examples include public participation, trustlessness, and censorship resistance.
		
\end{itemize}

Our categorization resulted in 21 technical primitives, 14 technical properties, 12 normative properties, 13 capabilities, and 15 use cases.
While we divide the concepts into these five categories, individual concepts are highly interconnected, both inter- and intra-category.
This provides credence to the notion that Blockchain technology overall is a cohesive whole, with each its component concepts serving a purpose in the overall technology.


This is the first stage of our methodology where research expertise directly influenced the results.
First, by its very nature drawing connections between concepts is subjective.
In most cases these connections were directly motivated by explicit references in the text, but in several cases we drew connections that we felt were implicit within the text.
Second, we identified several misconceptions that either shared no connections with the rest of the concepts or were obviously false (e.g., the assertion that cryptographic signatures provide confidentiality).
In both of these situations, our research notes kept track of what was explicitly supported by the analyzed data and what was the result of researcher interpretation.

\subsubsection{Stage 4---Theory Generation}
In the fourth and final stage, we used the concepts, categories, and connections derived from the first three stages of our grounded theory, along with our research notes and researcher expertises to derive several theories (i.e., research results from our analysis) regarding Blockchain technology.
First, we subdivided the technical properties into three categories that give a high-level description of what Blockchain technology is (see Section~\ref{sec:blockchain}).
Second, we extracted Blockchain technology's capabilities (Section~\ref{sec:capabilities}), related challenges (Section~\ref{sec:challenges}), and use cases (Section~\ref{sec:use-cases}).
Third, we found that there is a clean split between Blockchain technology's technological primitives and its normative properties (i.e., hype) (see Section~\ref{sec:normative}).

\subsection{Limitations / Research Artifacts}
Due to the nature of grounded theory, our analysis of the data represents one view on that data.
Different researchers coding the same data may have focused on different aspects leading to differences in categories, connections, and the theories they focused on.
To address this limitation, the analyzed documents as well as the generated codes, concepts, category graphs, and research notes will be available at [blinded for peer review].
We invite other researchers to review our work and to examine our collected documents for other interesting contributions.

\section{What is Blockchain Technology?}
\label{sec:blockchain}

\begin{figure*}
	\centering
	\includegraphics[page=2,scale=.75]{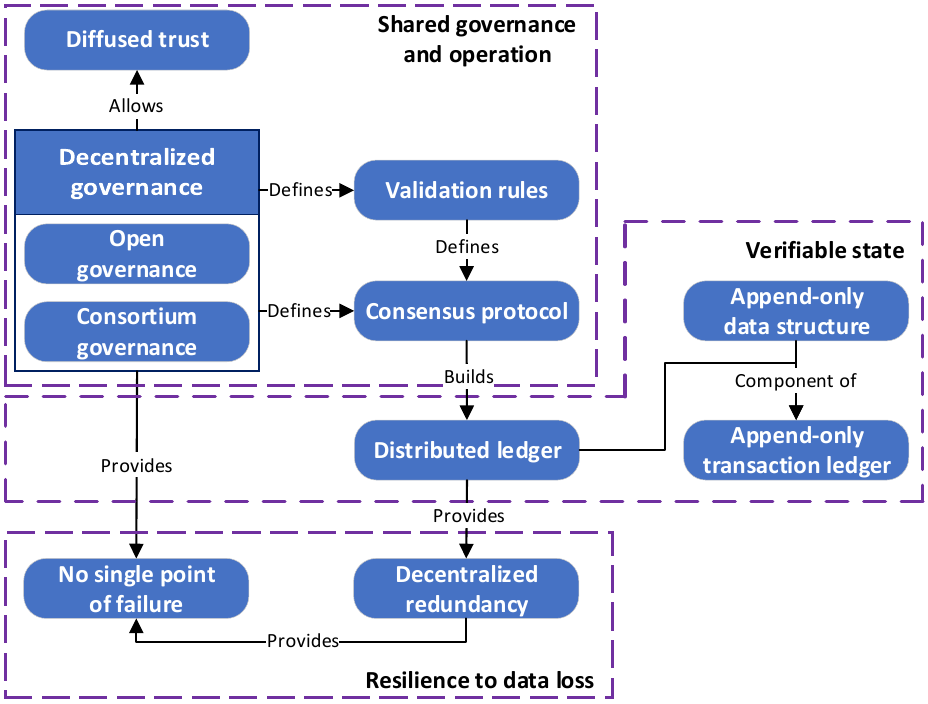}
	
	\caption{Technical Properties for Blockchain Technology}
	\label{fig:technical-properties}
\end{figure*}

Our literature analysis finds three key groupings of properties related to Blockchain technology (see Figure~\ref{fig:technical-properties}).\footnote{A version of this chart that shows the technical primitives that support the technical properties is given in the Appendix (see Figure~\ref{fig:technical-properties-full}).}
Most importantly, consensus is used to provide \textit{shared governance and operation}.
In support of shared governance and operation, other technical properties provide \textit{verifiable state} and {resilience to data loss}.
By themselves, these property groups are nothing new, but used together they define \emph{Blockchain technology}, or \emph{Blockchain} for short.
In Section~\ref{sec:distributed-comparison} we describe how other distributed technologies compare to Blockchain technology.

\subsection{Shared Governance and Operation}
\label{sec:sharedgov}


Blockchain technology was created to address the scenario in which a collection of parties----referred to hereafter as \emph{miners}---want to participate in a communal system but do not trust each other or any third party to operate the system singlehandedly.
By participating in both the governance and operation of the system (i.e., shared governance and operation), each miner can be assured that the system is operating correctly.
Even if some of the miners become compromised, the uncompromised miners retain the ability to detect malicious actions by the compromised miners and to prevent them from interfering with the correct operation of the system.
In this regard, Blockchain technology provides \emph{diffused trust} wherein it is not individual miners but rather the collective of all miners that is trusted.\footnote{This property has often been called ``trustlessness''; this is incorrect as trust still exists but has simply been diffused among multiple parties.}

Shared operation is enabled by the use of \emph{consensus protocols}, which are used by the miners to agree upon which operations---known as \emph{transactions}---will be executed by the system.
The consensus protocols allow miners to view transactions and validate that the system is operating appropriately, an important consideration as the miners do not trust that only valid transactions will be submitted to the consensus protocol.
Shared governance is provided by the ability of miners to configure their clients to only approve the transactions they believe are acceptable, effectively allowing them to vote for how the system should function.
If there is disagreement regarding system functionality, it is possible for a Blockchain-based system to split, resulting in the creation of competing systems that include only a subset of each other's transactions.
Usually these forks are temporary, with miners either choosing to all adopt the same rules, but it is possible for a fork to result in the permanent creation of two non-interoperable Blockchain systems (e.g., Bitcoin Classic and Bitcoin Cash).


Blockchain systems can be categorized based on who is allowed to act as a miner:\footnote{In our coding, we also identified the concept of private governance, which eschews the notion of miners. This concept is discussed later in Section~\ref{sec:private-blockchain}.}

\begin{itemize}
	\item \emph{Open governance (i.e., permissionless Blockchains).}
	Any party that is willing to participate in the consensus protocol is allowed to do so.
	These systems are susceptible to Sybil attacks, so it is necessary for them to use consensus protocols in which miners prove ownership and/or expenditure of some finite resource rather than relying on proofs of identity.
	Proof-of-work~\cite{DN93,back1997partial,NakamotoS8} (demonstrating ownership of computing resources) and proof-of-stake~\cite{Bano17,garay2018consensus} (demonstrating ownership of digital assets stored by the Blockchain system) are the most common methods.
	
	\item \emph{Consortium governance (i.e., permissioned Blockchains).}
	Only approved miners that can attest to their identity are allowed to participate in the consensus protocol.
	The starting set of approved miners is defined at system initialization.
	If this set never changes, it is known as a \emph{static consortium}.
	Alternatively, in an \emph{agile consortium} miners change over time, either based on the rules of the system (e.g., random selection) or through consensus by the existing miners.
	Because miners in a consortium have known identities, they can use Byzantine fault tolerance-based consensus protocols, which do not require the resource expenditure of the Sybil-resistant protocols used in open governance-based systems~\cite{Bano17,garay2018consensus}.		
\end{itemize}

For each type of governance, there is a need to incentivize correct participant behavior.
The first type of incentive is an \emph{intrinsic incentive}---i.e., miners maintain the system faithfully because they derive value from using it.
Next, \emph{on-chain incentives} exist when the Blockchain system provides direct benefits to miners for faithful execution (e.g., minting currency and giving it to the miners).
Finally, \emph{off-chain incentives} are any incentive that is not managed by the Blockchain system---for example, contractual obligations or individual reputation.
Importantly, off-chain incentives only apply to consortium governance as they inherently rely on knowing the identity of the miners.


\subsection{Verifiable State}
Miners adopt Blockchain technology because they want their trust to be rooted in the system---i.e., that the current state of the system accurately reflects the transactions that the consensus protocol allowed to execute in the past.
To enable this trust, Blockchain technology writes all transactions to a cryptographically-verified append-only ledger ~\cite{tamassia2003authenticated}, providing full system provenance and allowing miners (or outside parties) to audit the system's current state and past operations.
In many systems, including Bitcoin, this ledger is colloquially referred to as the ``blockchain'', but we avoid that term as it unnecessarily confusing to discuss both Blockchain (big-B) technology and the blockchain (little-b) data structure.

The first entry in the append-only ledger is known as the \emph{genesis block}.
The genesis block is responsible for specifying the initial parameters for the system.
Whenever a new transaction is approved by the miners, it is added to the ledger and cryptographically linked to one or more preceding transactions (or the genesis block for the first transaction)~\cite{bayer1993improving,haber1990time,haber1997secure}---for example, by signing the concatenation of the latest transaction and a hash of the transactions it is linked to.
The resulting data structure can be either linear (e.g., Bitcoin's hash chain) or branching (e.g., a Merkle tree or directed acyclic graph).
Regardless of the underlying structure, it is critical that all transactions are strictly ordered and that this ordering never changes after consensus is reached.


\subsection{Resilience to Data Loss}
If the ledger was only stored in a single location, deleting or modifying that data store could be detected by all parties, but there would be no guarantee that the data could be restored.
In Blockchain technology, the ledger is replicated among miners to address this single point of failure.
When data does need to be restored---for example, when an individual miner's ledger was corrupted---the replicated data can be verified to ensure that it correctly represents the system state.



Some Blockchain systems try to limit the amount of data any given miner needs to replicate by segmenting the data and assigning miners to handle governance and operations for only a subset of the system.
This is known as \emph{sharding}, with individual segments of the data known as \emph{shards}.
Sharding can drastically reduce the amount of data that miners need to store while also increasing the performance of the consensus protocol, which often scale based on the number of miners.
Still, sharding comes with the drawback that miners are no longer able to audit the system as a whole.
Additionally, by reducing the number of miners responsible for any given transaction, sharding also reduces the number of miners an adversary would need to compromise to attack that transaction.

\section{What are Blockchain Technology's Capabilities?}
\label{sec:capabilities}

\begin{figure*}
	\centering
	\includegraphics[page=4,scale=.75]{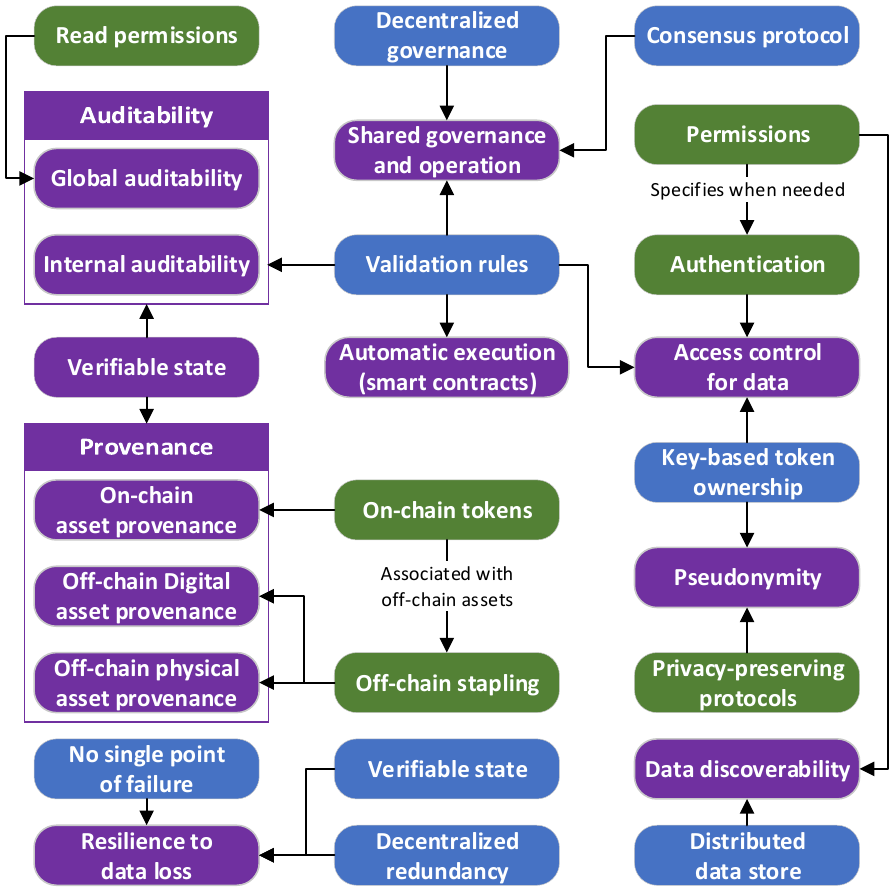}
	
	{\small Purple---capabilities, blue---technical properties, green---technical primitives. Arrows indicate that the destination depends on the source.}
	\caption{Capabilities for Blockchain Technology}
	\label{fig:Capabilities}
\end{figure*}

Capabilities define the high-level functionality that can be achieved by using Blockchain technology in a system's design.
Blockchain's three core capabilities were described in the preceding section: (1) shared governance and operation, (2) verifiable state, and (3) resilience to data loss.
In our coding, we identified 11 additional capabilities (see Figure~\ref{fig:Capabilities}).

\paragraph{Provenance and Auditability}
Blockchain systems provide a complete history of all transactions that were approved by the consensus process (i.e., full-system provenance).
This information can be used by the miners to audit the system and ensure that it has always followed the appropriate rules.
Additionally, this information can be used by non-miners to verify that the system is being governed and operated correctly.

If transactions are used to store information regarding digital or real-world resources, the resources must be \emph{stapled} to on-chain identifiers. Then the provenance information for the Blockchain system can also be used to provide audit information for those resources.
This can be used to track physical, off-chain assets (e.g., for supply chain management), digital, off-chain assets (e.g., copyrighted digital media), or digital, on-chain assets (e.g., cryptocurrencies or data files).
 
\paragraph{Access Control and Pseudonymity}
Data stored in a blockchain may have limitations regarding which users can use it as an input to a transaction or modify it as part of the operation of the transaction.
For example, a financial asset should only be a valid input to a transaction if the owner of that asset approves its use.
One approach to providing this functionality is storing access control lists (ACLs) in the blockchain and having the appropriate users prove their identity to the miners (e.g., using Kerberos or OAuth 2.0) as part of the transaction validation process.

More commonly, access control in a blockchain system is implemented cryptographically: data is associated with a public key when it is created and the ability to use or modify this data as part of transaction is granted only to users that can prove knowledge of the corresponding private key (e.g., by generating a signature over the transaction that validates with the public key attached to the data).
Ownership of the data can be expanded or transferred by associating it with a new public key.

On key benefit of access control using Blockchain is that the provenance of access control is automatically recorded.
This means that a full record of not only a user's permissions, but how they received those permissions, is stored.
This information can be used to automatically revoke permissions if it is discovered that a user was granted these permissions by a compromised account---for example, when a malicious insider grants inappropriate permissions to other insiders.

Key-based (as opposed to ACL-based) ownership of data has another advantage: it allows for pseudonymous ownership and use of data.
Still, this requires careful attention in the system design to use appropriate cryptographic techniques (e.g., zero-knowledge proofs, mix networks, or secure multi-party computation) to avoid linking real-world individuals to their keys and actions. This remains an open problem.

\paragraph{Automatic Execution (Smart Contracts)}
Blockchain transactions can also represent and store executable functions known as \emph{smart contracts}.
These smart contracts can be executed automatically in response to a function call in later transactions, with both the inputs and outputs of the function recorded within the calling transaction.
The smart contracts themselves are executed by the miners with outputs being verified through the consensus protocol.
The computational power of these scripts is determined by the system's rules, ranging from supporting only basic functionality (e.g., verifying a signature in Bitcoin) to providing Turing-complete functionality (e.g., Ethereum).

Smart contracts benefit from Blockchain technology's other capabilities (e.g., shared operation, auditability, and resilience).
For example, multiple miners execute and verify the output of a smart contract to help ensure that an adversary is unable to tamper with the result of a function.
Similarly, the ability to audit inputs and outputs can be used to attribute incorrect usage of a smart contract.
Still, smart contracts suffer from problems common to all programs (e.g., bugs, security flaws, complexity, or non-termination) and a failure to recognize this reality can lead to disastrous consequences.\footnote{This is best exemplified by the debate over ``code is law'' and the DAO attack: \url{https://www.coindesk.com/understanding-dao-hack-journalists/}.}

\paragraph{Data Discoverability}
If users are allowed to read any record in a Blockchain's distributed data store, then it is trivial to search for records of interest.
This capability is nothing more than what is provided by having a read-only data lake, but still it was frequently discussed in the literature we reviewed.

\section{Challenges for Blockchain Technology}
\label{sec:challenges}

Our analysis of the literature revealed several challenges that need to be considered when developing systems using Blockchain technology.
In this section, we describe these challenges and limitations.
In Section~\ref{sec:academic}, we survey academic research that is attempting to address some of the challenges.
Note that this section is focused on challenges facing Blockchain systems generally, not challenges facing specific applications such as Bitcoin or Ethereum).

\subsection{Scalability and Performance}
Many documents in our corpus point out that Blockchain technology's decentralized governance and operation incur significant overhead. The largest overheads incurred by Blockchain technology are (1) the need to run a consensus protocol before state can be updated, (2) the need to store the full system provenance, and (3) the need for each miner to store the ledger in its entirety. Any attempt to remove this replication comes at the cost of resilience that Blockchain-based systems often rely upon.

Additionally, most of today's open governance Blockchain systems are based on proof-of-work which bring additional challenges.
These proof-of-work schemes require users to acquire hardware and expend electricity to participate in consensus.
The real-world cost of these schemes can be tremendous---for example, it was estimated that as of April 2018 the energy consumed by Bitcoin miners alone was equivalent to the power usage of almost 5.5 million US households~\cite{Digiconomist}.

Another unintended consequence of proof-of-work is the centralization of mining power.  In order to reduce variance in their earnings, miners are incentivized to work together in large mining pools, pooling their computing power and sharing the profits among pool members. This phenomenon has emerged in all large-scale proof-of-work Blockchain systems\footnote{Currently almost 70\% of Bitcoin blocks are mined by the five largest mining pools~\cite{BlockchainInfoPools}}, and it is a problem because it inhibits decentralization~\cite{arxiv:GBERS18}.


\subsection{On-chain Correctness}
All executable code is subject to bugs---developer errors that can be taken advantage of to hijack program logic. This problem manifests in smart contracts, and when those contracts control the transference of valuable assets, the impact of a bug can be devastating. The immutability of a Blockchain's ledger exacerbates this challenge by impeding rollback of state changes, even those that are clearly malicious. This is because, by definition, any transactions on a blockchain upon which consensus is reached are considered legal---including ones due to buggy code and exploitations of such. If the miners decide to rollback the ledger to erase a mistaken transaction than confidence in the Blockchain system may be lost. Alternatively, if miners can't agree what to do about errant transactions, it could lead to a hard-fork in the Blockchain system.

Despite best efforts to eliminate mistakes in smart contract and transactions, a payment or asset transfer system must be able to reverse fraudulent or errant transactions. For example, if a user pays for a physical real-world good using a cryptocurrency but is then never given the purchased item. A new transaction reversing the effects of the disputed transaction could be added to the ledger, but decentralized governance makes arbitrating such a dispute difficult as there is no individual arbiter with the authority to determine which party is in the right when a dispute occurs.  Additionally, dispute resolution must be handled carefully to avoid introducing new vulnerabilities.  For example, several attacks were demonstrated against the Bitcoin refund mechanism~\cite{FC:MccShaHao16} necessitating further research to design secure refunds in Bitcoin~\cite{arxiv:AviSafSha18}.

\subsection{Off-chain stapling}
Another challenge for applications is ensuring consistency between on-chain state and the off-chain objects or state that it represents.  When Blockchain technology is used to track off-chain assets (physical or digital), those assets are typically represented on-chain by a digital identifier, or \textit{token}. When dealing with digital assets, correspondence between the asset and its token can generally be maintained by code; for example, a smart contract can track transference of ownership for a digital media license. For physical assets, however, maintaining this consistency is a challenge. Real-world processes must be employed to ensure that whenever an asset's state or ownership is modified, the corresponding token is updated. These processes are an obvious point of failure as they rely on correct enforcement by trusted parties. For example, a malicious entity could attach two tokens to one asset, two assets to one token, or issue tokens that have no backing asset (e.g. stocks in a naked short selling scenario). The end user must also be trusted, as they may be able to separate the token and sell it while keeping the asset, causing the token to be attached to an invalid asset (e.g., fake goods in luxury markets).

A related issue arises when Blockchain processes and smart contracts need to use off-chain inputs. For example, gambling contracts may determine which address to pay winnings to based on the result of a request to an off-chain oracle (e.g., sports scores, stock prices, weather forecasts, or other global events). If contract logic branches based on that response, the contract is no longer verifiable because auditors cannot confirm that the response received from the oracle at audit time is the same response received when the contract was executed. There are legitimate reasons why an oracle response might change with time, so this is really an inherent limitation of Blockchain technology: smart contracts cannot ``see'' external events.  Thus, additional mechanisms are needed to verify that these inputs cannot be forged.

\subsection{Security}
Due to its decentralized nature, Blockchain is potentially vulnerable to a number of security threats.  First, coordinated attacks by a majority (or often even a large minority) of the miners can reorder, remove, and change transactions from the ledger. Thus, it is critical that Blockchain applications provide the proper incentives to keep miners honest.  In particular, it is critical to design incentives such that the game theoretic behavior for selfish miners is to honestly maintain the state of the Blockchain system rather than to destroy it through forming such coalitions.

Additionally, Blockchain technology is vulnerable to traditional network attacks such as denial of service or partitioning.  Such attacks can aim to lower the number of participating miners or to fracture the network of miners to prevent consensus, lower the bar for 51\% attacks, or create an inconsistent state.

\subsection{Privacy and anonymity}
Another major challenge is how to protect the privacy of the users and data stored on a Blockchain ledger.  In basic implementations, data on the ledger is public in order to enable verification by all miners.  But this means that any sensitive data is inherently non-private. If confidentiality is needed, it will be necessary to either host a Blockchain system where only trusted entities can access it or by applying advanced cryptographic techniques that will allow miners to verify the correctness of encrypted transactions (e.g., multi-party computation, functional encryption). Still, the use of cryptography will limit the auditability and thus the ability to have meaningful shared governance.

Additionally, extreme care must be taken when trying to build an anonymous Blockchain system.
While many existing Blockchain systems provide a notion of ``pseudonymity'' in which users are identified by their cryptographic keys instead of by their names or social security numbers, it has been shown that this is not enough to provide true anonymity as attacks that correlate transactions by the same pseudonyms together with other data external to the Blockchain system can effectively deanonymize users~\cite{DBLP:journals/corr/abs-1708-04748}.

\subsection{Usability} 
For developers, development and analysis tools are critical to building secure applications in any domain. The availability of user-friendly developer tools varies significantly depending on the maturity of the target Blockchain platform. Some projects, like Ethereum have mature tools, while others have very little support. Many Blockchain platforms are currently geared towards expert users and lack the user experience-focused tools needed to allow for easier use by non-experts.

Another major challenge of some Blockchain systems is that the require users to store, manage, and secure cryptographic keys. However, this is known to be a significant impediment for most users ~\cite{uss:WhiTyg99}.  A survey by Eskandari et al.~\cite{EBSC15} outlines these challenges as well as potential solutions for managing keys for Bitcoin.  They discuss solutions such as password-protected and password-derived keys as well as offline and air-gapped storage of the keys.  But, as the authors state all of these solutions have their drawbacks.


\subsection{Legality and Regulation}
Our analysis revealed wide-spread concern with regulatory issues surrounding cryptocurrencies, Blockchain-based assets, and other Blockchain applications. It is important to note that regulation applies indirectly to technology, based on how the technology is used within a firm's operations. Therefore, there will be no direct regulation of Bitcoin, for example, but rather regulation of firms that use Bitcoin according to how they are using it. Consider the example of a Bitcoin exchange service that derives income from fees paid in Bitcoin: it will have to consider tax declarations as a business (\eg service taxes and capital gains), financial reporting as a money service business (\eg know your customer, anti-money laundering and anti-terrorist financing), generally acceptable accounting standards for audited financial reports (\eg reporting Bitcoin as an intangible asset on a balance sheet), and potentially additional regulation that applies to financial exchanges, banks, and/or custodians. In most countries, each of these already broad categories are administered by different government branches. Countries like the United States and Canada may require licensing or registration and have taken enforcement action against non-compliant firms.

An extreme case of regulation is prohibition of cryptocurrencies or Blockchain-assets. At the time of writing, the largest country to ban Bitcoin is Pakistan and the largest country to prohibit wide categories of cryptocurrency use is China.

\section{Blockchain Technology's Use Cases}
\label{sec:use-cases}

Within the literature we analyzed, there was significant discussion about potential applications for Blockchain technology.
We describe these use cases here.
In many cases, an argument could be made that these are not independent use cases. Some use cases are subsumed by others, or two use cases might have a common abstraction that make them effectively the same.
As authors, we attempt to categorize in a way that is useful to the reader trying to locate information about a particular use case. Before discussing the use cases themselves, we first discuss the general parameters of when a Blockchain is useful. 

\subsection{Financial Use Cases}

\paragraph{Electronic currencies and payments}


It is well-known that Blockchain technology can be used to build cryptocurrencies; Bitcoin is a working example of this.
Blockchain technology enable electronic transactions that are resilient even when large values are at stake.
Bitcoin has notable drawbacks that include scalability, performance and privacy.
There is ongoing research that demonstrates how Blockchain technology can be used to create payment systems that are low-latency and scalable, partially offline, confidential, and/or anonymous (see Section~\ref{sec:academic}).

\paragraph{Asset trading}

Real world financial markets provide the ability for the exchange of valuable assets. 
They tend to involve intermediaries like exchanges, brokers and dealers, depositories and custodians, and clearing and settlement entities. 
Blockchain-based assets---which are either intrinsically valuable or are a claim on an off-chain asset (material or digital)---can be transacted directly between participants, governed by smart contracts that can provide custodianship, and require less financial market infrastructure.
For tokens (i.e., digital identifiers) that represent something off-chain (\ie equity in a firm or a debt instrument), the issue of stapling must be addressed.
In most jurisdictions, financial markets are subject to government oversight making this area particularly encumbered by regulation.

\paragraph{Markets and auctions}

A central component of asset trading is the market itself---the coordination point for buyers and sellers to find each other, exchange assets, and provide price information to observers.
Auctions are a common mechanism for setting a fair price; this includes double-sided auctions like the order books in common use by financial exchanges. 
Decentralized markets and auction mechanisms can run on Blockchain technology.
The main challenge is the issuance of non-confidential transactions. 
An additional challenge is the potential for front-running by participants in the Blockchain network (in particular, miners) who learn of transactions before they are finalized. 

\paragraph{Insurance and futures}

A transaction can be thought of as a swap of one valuable asset for another. Swaps can be arranged for a future time.
Examples include agreeing to a price for a future purchase of oil, turning a variable interest rate into a fixed rate, and providing cash for collateral with the promise that the cash will be returned.
Further, this future swap might be contingent on an event happening. 
Examples include a derivative that is valuable if a stock price decreases, an insurance payout for a fire, or a payment that covers a loan default.
The primary challenge of transactions of this the risk that the counterparty will not fulfill their future obligations. While Blockchain technologies can reduce some types of trust, it cannot easily solve counterparty risk. 
It can offer transparency which can be used to build reputation and contracts can be designed to hold digital currency or assets as collateral and disperse them if Blockchain-based conditions are met.
A second challenge for many of these use cases is reporting real world events (a fire, a change in the price of a stock, or mortgage default) to the Blockchain in a trustworthy fashion.
However, this is not an issue for events that are Blockchain-based to begin with.
This demonstrates a key point: deploying several complimentary use cases on the same Blockchain enables complex interactions.

\paragraph{Penalties, remedies, and sanctions}

In common parlance, code running on a Blockchain is called a smart contract.\footnote{In particular, the success of Ethereum contributed to this, although that project now prefers the term `decentalized app' or dapp for short.} Legal contracts will often anticipate potential future breaches and offer a set of penalties or remedies.
With Blockchain technology, a set of remedies could be programmed into the contract assuming both the triggering event and the resulting action are Blockchain-based (or the real-world/Blockchain gap is bridged by a trustworthy entity).
If the contract is legally well-formed, with identified counterparties in a clear jurisdiction, the remedies can be thought of as a set of reasonable default actions that can avoid, but do not preclude, expensive litigation. 


\subsection{Data Storage and Sharing Use Cases}

\paragraph{Asset tracking} 

Blockchain technology can be used for tracking material assets that are globally distributed, valuable, and whose provenance is of interest.
This includes standalone items like artwork and diamonds, certified goods like food and luxury items, dispersed items like fleets of vehicles, and packages being shipped over long distances, which will change hands many times in the process. 
It also includes the individual components of complex assembled devices, where the parts originate from different firms. 
For heavily regulated industries, like airlines, and for military/intelligence applications, it is important to establish the source of each part that has been used, as well as a maintenance history.
While assets are already tracked in digital databases, there is no common database shared by each participant in the supply chain.  

A Blockchain sidesteps the political problem of who should host such a shared database when the candidates are competing firms and government agencies from different jurisdictions. 
Blockchain technology provides a common environment where no single firm has the elevated power and control of running a widely-used database. 
The main integrity challenge is the stapling issue: specifying how material assets are assigned a tracking token on the Blockchain in a trustworthy manner. 
A second challenge is the lack of confidentiality Blockchain technologies offers by default when the data is proprietary and tied to profitable business practices.
Finally, a third challenge is getting agreement on the technology to be used (this is being explored through business consortiums.\footnote{\eg Blockchain in Transport Alliance}).

\paragraph{Identity and key management}

Identities, along with cryptographic attestations about properties for those identities (e.g., over 18 years of age, has a driver's license), can be written to the Blockchain.
These identities and attestations can then be used by other systems to support their access control policies.
Importantly, this identity information comes with full provenance. This could be useful in determining suspicious activity (e.g., having an age that is not increasing linearly).
This could also be a quicker and more performant way of establishing identity than the current certificate authority system.

\paragraph{Multi-organization data sharing}

Asset tracking and identity tracking are both examples of sharing data across organizations, and Blockchain technology contributes a common environment.
The use cases in this category share challenges: Blockchain technology can specify write access policies to data stored on the Blockchain, but it will not provide any default support for restricting read access; confidentiality has to be an additional layer.
Further, the integrity of confidential data cannot be validated by nodes in the network without some minimal disclosure of what the data is or that it satisfies the relevant restrictions.
Blockchain can also serve as a secondary component in these systems, where capabilities are issued and transferred as if they were financial assets on a Blockchain. Proving ownership over a capability is done off-chain to a traditional enforcement server, which enables the correct permissions.

\paragraph{Tamper-resistant record storage}
The append-only ledger is used to store documents, including the history of changes to these documents.
This use case is best suited for records that are highly valuable (such as certificates, government licenses), have a small data size, and are publicly available (as they will be replicated by all miners).
If large and/or confidential documents need to be stored, then the Blockchain can store binding/hiding commitments for the documents, while the documents themselves are stored in another system with lower overhead.
Blockchain technology could be used to timestamp documents. Still, timestamping generally does not rely on any of Blockchain technology's key properties, and so Blockchain technology is likely overkill for this application.




\subsection{Other use cases}

\paragraph{Voting}
Electronic voting is a challenging problem that is often asserted to benefit from Blockchain technology's properties.
Shared governance could be used to ensure that multiple parties (the government, non-governmental organizations, international watchdogs) can all work together to ensure that an election is legitimate.
Audibility is important in providing evidence to the electorate that the election was fair.
Finally, the resilience of Blockchain technology is important in preventing cyberattacks against the voting system.
However, voting on a Blockchain has many challenges to solve: Blockchains offer no inherent support for secret ballots, electronic votes can be changed by the device from which they are submitted (undetectably if a secret ballot is achieved), cryptographic keys could be sold to vote buyers, and key recovery mechanisms would need to be established. 

\paragraph{Gambling and Games}

Examining the most active Bitcoin scripts and Ethereum decentralized applications shows that gambling is quite popular. Players can audit the game to ensure that execution is fair, and the system can operate its own cryptocurrency to handle the finances (including holding the money in escrow to prevent losing parties from aborting before paying). This use case is best suited to gambling games that do not require randomness, private state, or knowledge of off-blockchain events. For the set of residual games,\footnote{Currently, the most active Ethereum game is called Fomo3D: users pay to reset a 30 second countdown timer and if it ever reaches zero, the last user to pay wins all the money collected.} Blockchain is an ideal platform. For the other types of games, new layers of technology would have be added on a Blockchain. Data feeds (called oracles) of either randomness or real world event outcomes requires additional trust and introduces finality risk, while confidential user state require additional cryptography.  

\paragraph{IoT and smart property}
IoT devices occasionally have the need to collectively make decisions.
In these cases, Blockchain technology can provide a technological platform for making these collective decisions in an auditable fashion.
This auditability is especially important as IoT devices are notoriously untrustworthy due to insufficient security, and the ability to later audit and analyze their actions is invaluable.
Additionally, the replication inherent to Blockchain technology means that even if a subset of the IoT devices is lost (e.g., destroyed sensors in a storm), it can still be possible to record the entire provenance of all devices.



\section{Is Blockchain the Right Solution?}
\label{sec:distributed-comparison}

Blockchain technology provides a unique set of capabilities that might be better suited for a system design than competing database technologies or distributed systems.
However it does come with a relatively high overhead: replicating all past and present data and operations on the data at every node of the network in an auditable ledger.
Based on our results and experience we recommend the use of the following questions to determine if Blockchain technology would be a good fit for a specific project (see Section~\ref{sec:sharedgov} for definitions).

\begin{enumerate}
	\item Does the system require shared governance?
	\item Does the system require shared operation?
\end{enumerate}

If the answer to both questions is no, then Blockchain's consensus protocol is likely unnecessary overhead. If the answer to both questions is yes, then Blockchain technology is likely a good fit. This is due to the fact that meaningful shared governance \emph{and} operation requires miners to audit the operations of others and to be able to recover data that a malicious miner might try to delete (questions 3 and 4 below, respectively). If only shared governance or shared operation is needed, then the following two questions can be used to determine if the auditable ledger and replication, respectively, justifying the use of Blockchain technology if both are needed:

\begin{enumerate}[start=3]
	\item Is it necessary to audit the system's provenance?
	\item Is it necessary to prevent malicious data deletion?
\end{enumerate}

\subsection{Relationship to other distributed systems}
Blockchain technology fits within the broader family of distributed systems.
At the highest level, Blockchain technology is a type of decentralized database.
To help readers situate Blockchain technology within this greater ecosystem we have created a taxonomy and a flowchart based on that taxonomy (see Figure~\ref{fig:blockchainFlowchart}).

\begin{figure*}
	\centering
	\includegraphics[width=\textwidth]{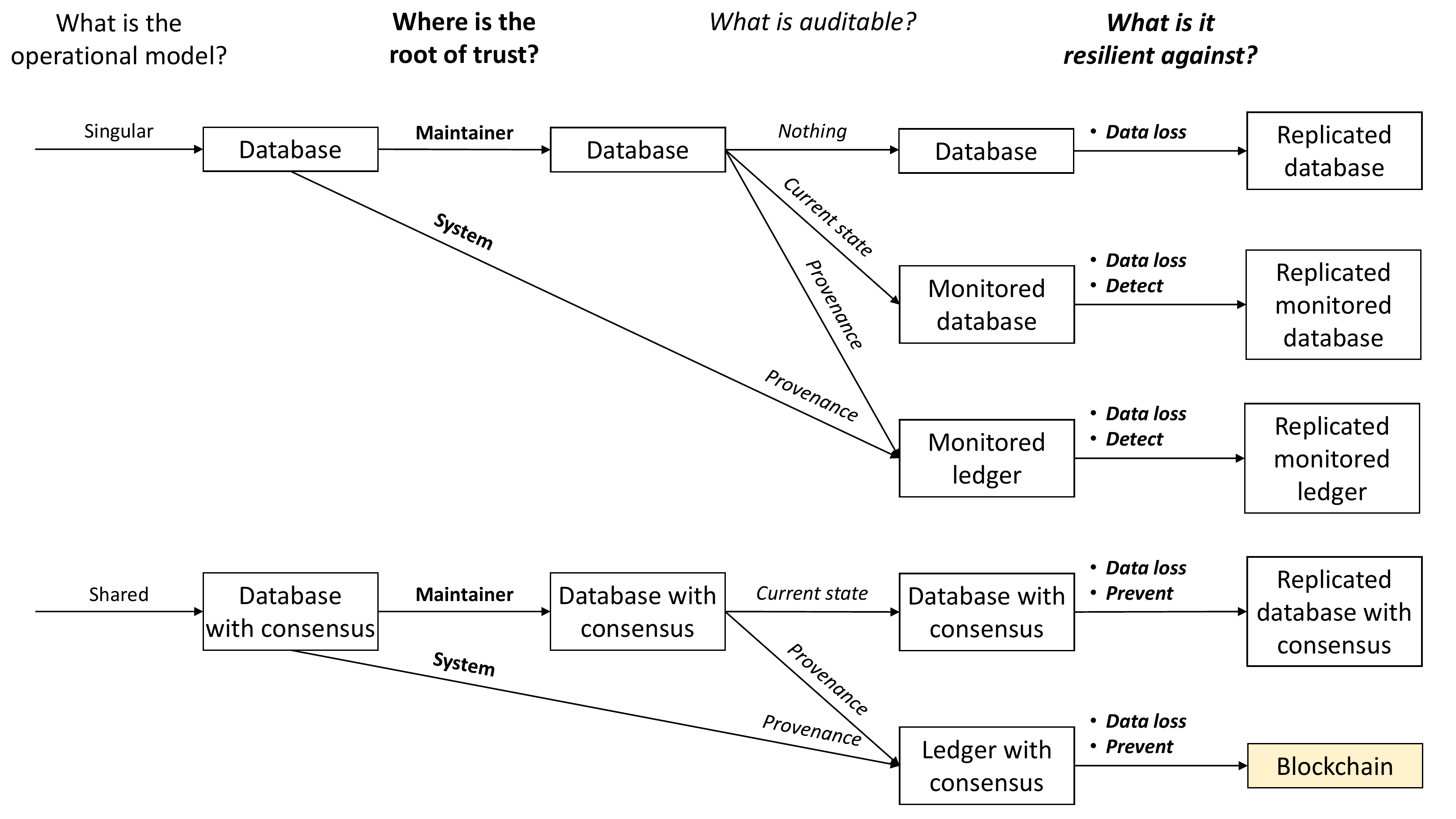}
	\caption{Comparing decentralized databases}
	\label{fig:blockchainFlowchart}
\end{figure*}

The first property in our taxonomy considers who has the authority to manage and update the database: \emph{what is the operation model?} In a singularly governed database (``Singular''), a single entity performs these tasks. Alternatively, the system can use a consensus protocol to allow for shared governance (``Shared'').

Next, we consider the security model by asking \emph{where is the root of trust?}
This refers to the entity or entities that must behave honestly in order for the system to be secure.
In typical database systems, trust is rooted in the maintainer (``Maintainer'')---for example, using AWS cloud storage requires that you trust Amazon.
Alternatively, trust can be rooted in the design of the system itself (``System''), though this is only possible if the system stores sufficient provenance for it to be audited to confirm that the system is functioning as intended.

The next question is \emph{what is auditable?}
In the worst case, nothing is auditable (``Nothing'').
Systems can use an authenticated data structure~\cite{tamassia2003authenticated} to ensure that their current state can be audited (``Current state'').
If the state also contains a history of the system (e.g., a ledger), then the use of an authenticated data structure allows for the provenance of the system to also be audited (``Provenance'').
In both cases, it is necessary that these databases be monitored to ensure that they never enter an invalid state, even temporarily.
In the case of shared operation, the operating entities act as monitors of the current sate during the consensus protocol.

Finally, we can classify systems by asking \emph{what is it resilient against?}
In particular, we considered with three resiliency properties---(1) is it resilient to accidental data loss (``Data loss''), (2) is it possible to detect that data has been malicious altered (``Detect''), (3) and is it possible to prevent malicious updates (``Prevent'').
Replication is a simple solution to prevent against accidental data loss, but by itself it fails to prevent malicious data loss as the malicious changes to the database will also be replicated.
In singular operation, external monitors can help detect malicious data changes, but can only do so after the data has been lost.
In shared operation, the monitors are the operators and malicious deletions and modifications will not be replicated, preventing them from effecting the overall system.

\section{Discussion}
In this section we discuss the remaining results from our grounded theory work.


\subsection{Normative properties}
\label{sec:normative}

\begin{figure*}
	\centering
	\includegraphics[page=3,scale=.75]{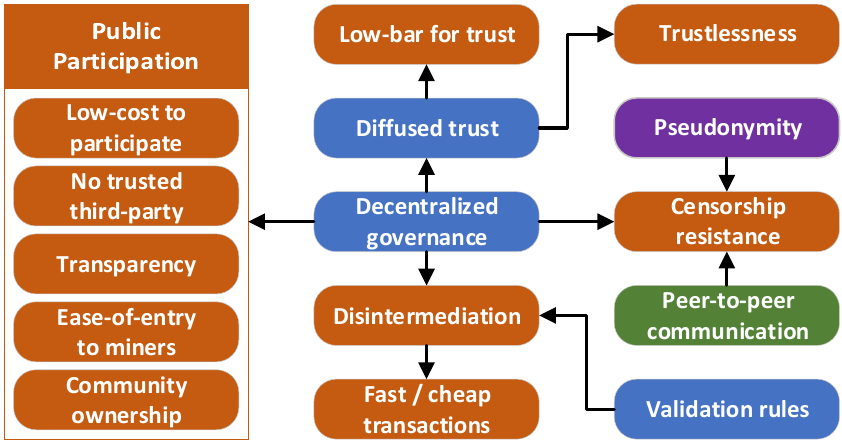}
	
	{\small Orange---normative properties, purple---capabilities, blue---technical properties, green---technical primitives. Arrows indicate that the destination depends on the source.}
	\caption{Normative Properties for Blockchain Technology}
	\label{fig:normative-properties}
\end{figure*}

Within the literature we analyzed, there were a set of properties that were not technical properties directly provided by Blockchain technology, but rather expressed desired properties for systems built using Blockchain technology (\ie normative properties).
These properties are shown in Figure~\ref{fig:normative-properties}.

The majority of the normative properties focused around the notion of using Blockchain technology to allow for public participation.
While public participation is certainly possible with Blockchain technology, as there are many examples of systems built using Blockchain technology that fail to achieve these properties.
For example, while Bitcoin initially had a low-cost to participate allowing easy-of-entry for miners and community ownership, that is no longer the case as any meaningful participation requires the purchase of a large amount of specialized hardware and the expenditure of a significant amount of electricity.

Similarly, the normative properties of low-bar for trust, disintermediation, no trusted third-parties, censorship resistance, and fast/cheap transaction all require extremely careful system design to achieve and are not guaranteed by the use of Blockchain technology.
In practice, these properties are often difficult to guarantee for any system, and this is no different for Blockchain technology.
For example, Bitcoin ultimately requires trusted third parties (e.g., exchanges, retailers who accept Bitcoin) to allow the currency to be useful for real-world application.
Also, fast/cheap transactions usually only exist because Blockchain based systems are not yet regulated like non-Blockchain systems.

When reading documents from industry, normative and technical properties are often intermingled with each other.
The injection of ideology into a technical field causes confusion and suboptimal design choices, not to mention muddying discussion and preventing clarity.
Interestingly, in the concept graph generated by our methodology, the technical and normative properties were cleanly separated.
No capabilities have dependencies on normative properties and removing them from the graph does not lessen the value of the graph as an exploration of technical concepts.
The fact that this separation occurred naturally provides evidence that grounded theory accomplished our research goals and was a good choice for addressing this corpus of data.


\subsection{Private Governance}
\label{sec:private-blockchain}
In our survey of the industrial literature, we encountered several proposals for Blockchain-as-a-Service.
While such services often use true Blockchain protocols, they also use a centralized governance model where all miners are controlled by a single entity (i.e., private governance).

We do not classify such systems as Blockchain technology.
First, these systems do not neither shared governance nor shared operation, which we identify as the key component of Blockchain technology.
Second, the entity operating the system still represents a single-point of failure.
While the miners within the operating organization might be run on a distributed infrastructure, there is still a high chance that a compromise in the operating organization would lead to a compromise in the Blockchain system.
Third, there is nothing that prevents the governing party from deleting or modifying data; even if such changes could be detected, the data itself is not replicated outside the organization and would be lost.
This is not to say that such systems lack value---such an evaluation is beyond the scope of this paper---but rather we believe that these types of systems are distinct from the more decentralized Blockchain technology.

\subsection{Lack of Privacy and Data Discoverability}
In the literature we found a common misconception that Blockchain technology inherently provided confidentiality for information stored within it.
In fact, the opposite is true: all transactions are visible to all miners, and this is necessary for miners to validate transactions.
The global visibility was identified by some as a capability (i.e., data discoverability) that allowed a Blockchain to act as a data lake.
While there were some valid applications of Blockchain as a data lake, in most cases we found that proposed data lake applications did not need all of Blockchain technology's capabilities and that a simpler solution would have sufficed.
It may be possible to add confidentiality to Blockchain technology, but care must be taken to ensure that this confidentiality does not preclude miners' ability to validate and audit the system.
This remains an open research problem.

\subsection{Ideology, Hype, and Ulterior Motives}
Many proponents of Blockchain technology believe that it has the capability to massively disrupt how society operates, or at least to rapidly overtake legacy solutions in many significant industries. This belief is hyperbole, as though Blockchain technology has many valid uses, it has not, nor is it likely to achieve this Utopian vision. This ideology and hype causes problems: for example, frequent emotionally-charged schisms within Blockchain advocate and developer communities---especially those affiliated with Bitcoin. This turmoil prevents level-headed scientific discourse and wastes developer resources. It can also tangibly affect the stability of a Blockchain system by causing a fork in which two independent chains emerge to used and maintained by different groups, further dividing resources.

With that said, Blockchain's disruptive power has certainly been demonstrated in the financial sector, so it clearly has promise. Several factors have made this sector an attractive target for disruption, perhaps none more so than the opportunity for massive profit. This motive has had benefits for Blockchain, especially in accelerating the pace of technological development. However, it has also created perverse incentives to reinforce hype and ideology. Hype can attract investors and inflate valuations, and dogmatic ideology is a proven marketing and recruitment strategy for financial scammers. These problems inhibit the advancement of Blockchain technology.

\subsection{Reputation for illicit uses}
Due to the prominence of Bitcoin, many people are familiar with Blockchain first and foremost as the technology underlying the cryptocurrency and therefore the reputations of the two are intertwined. The fact that Bitcoin is designed to avoid banks and central authorities in general, combined with its well-known history of illicit uses, somewhat poisons the well for Blockchain as a whole. Along with the causes listed above (ideology, hype, and ulterior motives), this contributes to the difficulty of discussing and considering Blockchain technology with precision and objectivity. It may also have impeded or delayed its acceptance by organizations unwilling to associate themselves with the Bitcoin's poor reputation.

\section{Survey of Academic Research on Challenges}
\label{sec:academic}
As part of the grounded theory analysis, the data revealed several open research challenges related to Blockchain technology.
In regards to these challenges, we surveyed the academic literature to identify what researchers are doing to address these challenges.
In this section, we give a brief over of the results of our literature survey.
This is not intended to be an exhaustive review of the space, but rather to highlight areas of substantial work.

\subsection{Blockchain scalability}

\subsubsection{Power consumption and centralization of mining}

To reduce the power consumption of open governance Blockchain systems that have traditionally relied on a proof-of-work-based consensus protocols, there have been several proposals to turn to other consensus mechanisms (for a more complete survey of this space see~\cite{Bano17}). The most popular of these proposals is proof of stake consensus, where parties' contribution to the consensus protocol is proportional to the total amount of stake they own in the system rather than the amount of work that they do. Today's proof of stake protocols (e.g.~\cite{FC:BenGabMiz16,eprint:BenPasShi16,CRYPTO:KRDO17,SOSP:GHMVZ17}) vary significantly in their model, assumptions, and performance guarantees.  Other suggestions for avoiding proofs of work include proof of space~\cite{CRYPTO:DFKP15, eprint:PPKAFG15} where miners use storage instead of computation, and proof of elapsed time~\cite{SSS:CXSGLS17} where trusted hardware (i.e., Intel SGX) is used in place of proofs of work.  It is not clear at this point which of these solution will be best suited for different Blockchain deployments.  

Another unintended consequence of proof-of-work consensus is the centralization of mining power.  In order to reduce variance in their earnings, miners are incentivized to work together in large mining pools, pooling their computing power and sharing the profits among pool members.  Currently, almost 70\% of Bitcoin blocks are mined by the five largest mining pools \cite{BlockchainInfoPools} significantly limiting the actual decentralization achieved by Bitcoin~\cite{arxiv:GBERS18}. Of course, these pools are disincentivized to damage trust in Bitcoin (and thus reduce its value and their profits) by abusing their power to censor transactions or violate rules in other ways. But this centralization undoubtedly runs counter to the goals of open governance and may violate security notions that depend on decentralization.  One possible solution~\cite{CCS:MKKS15} is to discourage mining pool formation by making it impossible to enforce cooperation between the members.

\subsubsection{Increasing transaction rates}
Another challenge to the scalability of open governance Blockchain solutions is the increasing number of transactions.
Current systems often have rather long wait times before a transaction can be confirmed  (e.g., Bitcoin can take several hours to confirm a transaction~\cite{BlockchainInfoTransactionConfTime}).  This makes these solutions less than ideal when immediate transactions are needed, such as when purchasing physical goods.

A couple of different approaches have been proposed to deal with this issue.  First, a number of hybrid consensus algorithms (e.g.,~\cite{SOSP:GHMVZ17,OPODIS:AMNRS17,DISC:PasShi17,EC:PasShi18,NSDI:EGSR16}) aim to reduce transaction approval times through reducing or eliminating forks. Most of these work by using ``proof-of'' style protocols to elect a committee or a leader who then uses traditional byzantine fault-tolerant consensus. A second approach for improving transaction rates, especially for financial transactions, is to make use of payment-channel networks. Such networks set up pairwise channels between parties to allow transactions on these channels to occur ``off-chain'', i.e., without being recorded on the ledger; the ledger is then only used for conflict resolution.  Many different flavors of payment-channel networks achieving various properties have been proposed (e.g.~\cite{PooDry16, NDSS:HABSG17,CCS:KhaGer17,SYSTOR:LNEKPS18,CCS:MMKMR17,CCS:GreMie17}) and several such as the Lightning network~\cite{PooDry16} are in active development for financial transactions on top of Bitcoin.  

\subsubsection{Handling increased transaction volume}
Another scalability challenge for popular Blockchain-based systems, such as Bitcoin and Ethereum, is the sheer volume of transactions that are being added to the ledger.  As more and more third-party services start to use these systems to store and execute their transactions, these systems have to verify and store transactions for a variety of unrelated operations. This can cause the storage and verification work required of miners to become prohibitively expensive.

Proposed solutions for this problem include sharding (e.g.~\cite{CCS:LNZBGS16, FC:GenRenSir17}) to partition transactions based on the transaction type or service.
This allows different sets of miners to verify transactions for different services thus reducing the amount of verification work each miner must do.  Another more radical approach to deal with this challenge has been to move away from the ``chain'' view of Blockchain technology.  Instead, several proposals (e.g., ~\cite{ePrint:SomLewZoh16,eprint:SomZoh18,IOTA}) propose to organize transactions into a directed-acyclic graph (DAG) where later transactions can vote on the validity of earlier transactions, allowing transactions to be approved before global consensus is achieved.
\subsection{Smart contract correctness and dispute resolution}

Three different directions have been proposed for improving the correctness and security of smart contracts:  Education and tools to help developers write smart contracts, tools for evaluating correctness and security of existing smart contracts, and formal modeling and formal verification of smart contracts. Along the education path, researchers organized a class on developing smart contracts cataloging common mistakes and misunderstandings~\cite{FC:DAKMS16}. Additionally, tools have been developed to simplify development of \emph{private} smart contracts~\cite{SP:KMSWP16}. For evaluation of existing smart contracts, multiple tools using symbolic execution~\cite{CCS:LCOSH16}, machine learning~\cite{arxiv:Huang18}, and static analysis~\cite{CCS:BDFGGK+16,NDSS:KGDS18} have been developed for detecting bugs and vulnerabilities. Finally, some efforts to support development of formally verified smart contracts is underway; for example, the Ethereum Virtual Machine (EVM) has been fully defined for interactive theorem provers~\cite{Hirai17}, which are essential tools for building formally verified software of any kind. Magazzeni et al.~\cite{Magazzeni17} have laid out a research agenda identifying further groundwork that must be conducted to support formal verification of smart contracts.

\subsection{Key Management}
Some recent work has also looked into cryptographic solutions to protect users' keys.  Techniques based on secure multi-party computation (MPC)~\cite{CCS:LinNof18,C:Lindell17} allow transactions to be signed without any party ever having access to the secret key.  Alternatively, the classic technique of threshold signatures (e.g.~\cite{PKC:Boldyreva03,EC:GJKR96,EC:Shoup00a}) allow users to split their keys into many pieces such that a large number of them must be compromised in order to ``steal'' a user's key.  However, much work remains to secure all the cryptographic keys inherent in real-world blockchain deployments.

\subsection{Regulation}
As applications of blockchain technology proliferate, they have drawn significant attention from the regulatory bodies around the world.  In the settings of cryptocurrencies, a number of concerns such as the prevalence of black-market transactions, tax evasion, money laundering, and terrorist financing have drawn calls to regulate how such cryptocurrencies can be used.  An excellent review article by Kiviat~\cite{Kiviat15} outlines some of the issues that arise in regulating blockchain transactions.  Additionally, as blockchain applications move to greater support of smart contracts, researchers have shown that criminal smart contracts are easily implementable in today's smart contract platforms~\cite{CCS:JueKosShi16} requiring regulation to avoid such criminal uses of blockchain.  

%


\section{Related Work}
\label{sec:related-works}

Different aspects of the Blockchain landscape have been systemized in past work, however our approach provides a unique and complimentary perspective. An early comprehensive work is Bonneau \etal's cryptocurrency systemization of knowledge~\cite{BMC+15}, which advocates for research on Bitcoin, merges disparate non-academic sources of information, and evaluates extensions that begin to tread beyond currency. We share a common approach of bringing non-academic work into an academic light, however we take the broader focus of Blockchain applications beyond cryptocurrencies as our starting point, we take greater effort at applying a thorough methodology for the evaluation of non-academic work, and we draw from a different body of knowledge (\ie from industry practitioners instead of the developer community).

More recently, W{\"u}st and Gervais develop a flow chart to answer the question: ``do you need a Blockchain''~\cite{Wust17}, and they evaluate several use-cases (that overlap with the ones we extract) with it. The authors use an approach based on domain knowledge and technical expertise; we purposely seek to minimize our own researcher bias to ascertain how non-experts understand the technology. Their flow chart is compared favorably to 30 similar charts appearing in industry whitepapers (that overlap with our dataset) and blog posts studied by Koens and Poll~\cite{koens2018blockchain}. This minimizes the novelty of the flowchart we develop in Figure~\ref{fig:blockchainFlowchart}, but it is a minor contribution of this work. 

Narayanan and Clark describe the ``academic pedigree'' of Bitcoin's core technical innovations, repudiating the common belief that Bitcoin was a radical departure from existing research~\cite{Narayanan17}. The authors touch lightly on the public (mis)understanding of Blockchain, highlighting some key misconceptions, however our work explores this thoroughly.


Several surveys deal with specific technical topics including consensus and scalability~\cite{Gervais16,Croman16,Bano17,garay2018consensus}, security vulnerabilities~\cite{Conti17}, and privacy issues~\cite{Henry18}. Our work has the broader focus of situating Blockchain's general capabilities in potential industry use-cases.



\section{Summary}
In this paper we answer four common questions regarding Blockchain technology: (1) what exactly is Blockchain technology, (2) what capabilities does it provide, and (3) what are good applications for Blockchain technology, and (4) how does it relate to other approache distributed technologies (e.g., distributed databases).
We accomplish this goal by using grounded theory to analyze a large corpus of data produced by industry.
Ultimately, Blockchain technology is neither a panacea nor worthless.
Instead it is a useful tool in a system developer's toolkit that can be applied when its overhead (consensus and replicated full system provenance) is justified by the system's needs (shared governance and operation, verifiable state, and resilience to data loss).
Even though Blockchain technology does not solve all the problems that its proponents claim it does, we believe that is a meaningful technology that will continue to be used in industry and is deserving of some attention by industry.


\bibliographystyle{IEEEtran}
\bibliography{bib/bitcoin}


\clearpage
\onecolumn
\appendix

\subsection{Technical Properties Full Diagram}


\begin{figure}[h]
	\centering
	\includegraphics[page=1,width=\columnwidth]{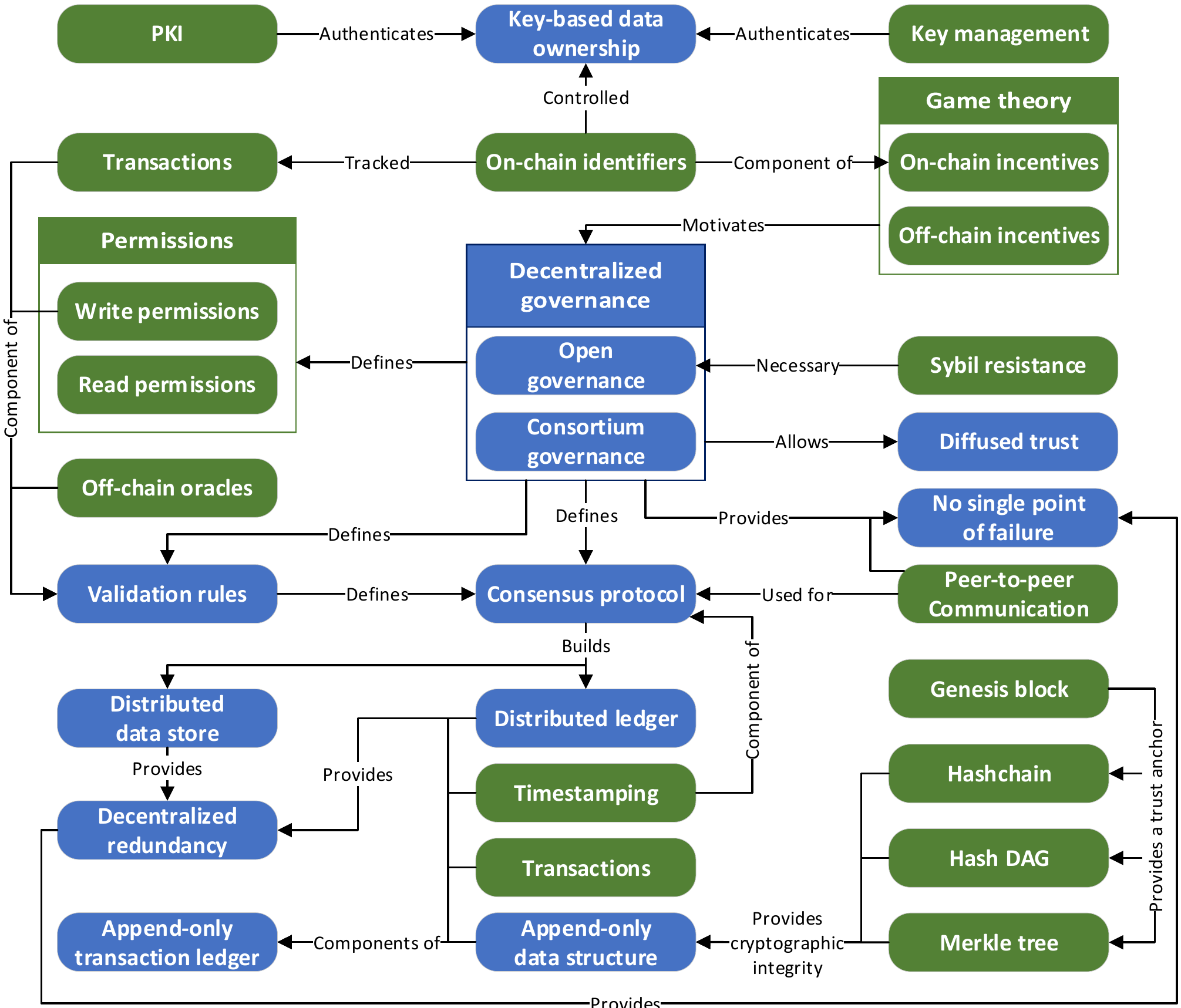}
	
	{\small Blue---technical properties, green---technical primitives}
	\caption{Technical Properties and Primitives}
	\label{fig:technical-properties-full}
\end{figure}


\end{document}